# Estimating Tipping Points in Feedback-Driven Financial Networks

Zvonko Kostanjčar[1], Stjepan Begušić[1], H. E. Stanley[2], and Boris Podobnik[2−5]

September 16, 2015

[1]University of Zagreb, Faculty of Electrical Engineering and Computing, Zagreb, Croatia.
[2]Boston University, Center for Polymer Studies and Department of Physics, Boston, USA.
[3]University of Rijeka, Faculty of Civil Engineering, Rijeka, Croatia.
[4]University of Ljubljana, Faculty of Economics, Ljubljana, Slovenia.
[5]Zagreb School of Economics and Management, Zagreb, Croatia.

**Abstract**

Much research has been conducted arguing that tipping points at which complex systems experience phase transitions are difficult to identify. [1, 2, 3, 4, 5, 6, 7] To test the existence of tipping points in financial markets, based on the alternating offer strategic model we propose a network of bargaining agents[8, 9, 10, 11, 12, 13, 14] who mutually either cooperate or compete,[18, 15, 16, 17] where the feedback mechanism[19, 20] between trading and price dynamics is driven by an external "hidden" variable $R$ that quantifies the degree of market overpricing. Due to the feedback mechanism, $R$ fluctuates and oscillates over time, and thus periods when the market is underpriced and overpriced occur repeatedly. As the market becomes overpriced, bubbles are created that ultimately burst in a market crash. The probability that the index will drop in the next year exhibits a strong hysteresis behavior from which we calculate the tipping point. The probability distribution function of $R$ has a bimodal shape characteristic of small systems near the tipping point. By examining the S&P500 index we illustrate the applicability of the model and demonstate that the financial data exhibits a hysteresis and a tipping point that agree with the model predictions. We report a cointegration between the returns of the S&P 500 index and its intrinsic value.

# 1 Introduction

.

Although much network science research has focused on how network collapse occurs when certain internal parameters approach their tipping points,[21] there is a broad class of real-world complex networks in which the dynamics are driven by "hidden" *external* variables[22, 23, 24] in the form of feedback mechanisms. The tipping points at which these networks collapse have not yet been adequately understood. In a buyer/seller financial network the trading dynamics are strongly affected by trader perceptions that the market is over-priced or underpriced. In a landlord/renter network a hidden variable is the ratio between the average apartment price and the average renter income. In a university/student network a hidden variable is the ratio between the average tuition and the average family income. In each of these examples, increasing the ratio increases the probability of network collapse, and our goal is to quantify the ratio that causes networks to collapse.

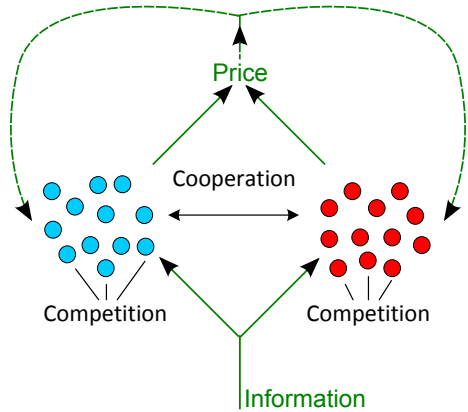

Figure 1: Competition, cooperation, and feedback mechanism between trading process and price dynamics in a coupled network model. Inflow of information determines intrinsic price, a genuine, fundamental information about an asset. The ratio between market price and intrinsic price affects the trading dynamics which in turn change the trading pattern.

Here we focus on financial trading markets where the trading decisions made by the market players are influenced by their expectations about the future. With the passing of time they learn whether their expectations about future market behavior were accurate. Overly-optimistic predictions in particular are very time-limited and are often followed by adjustments in the market price so abrupt that they cause the market to collapse.

Here we propose a network of bargaining agents and demonstrate how the degree of market overpricing through a feedback mechanism between trading and price dynamics induces further market overpricing, market bubbles, and utimately market collapse[27, 28, 29, 25, 26]. According to Scheinkman a bubble is a period in which prices exceed fundamental value.[30] We find that a well-known US financial index, the S&P500 index, exhibits a hysteresis behavior and a market-collapse tipping point that confirms the predictions of our network model.

## 2 Complex network market model

### 2.1 Initial Networks.

Because most human activity is limited by the finite availability of resources, individuals are compelled to bargain over the division of those resources.[8, 18] Bargaining has been at the core of trade from the earliest recorded human history when, prior to the introduction of currency, goods and services were bartered. Today bargaining is ubiquitous and ranges from haggling for food items in certain cultures to negotiations between large international business firms.[31, 23, 29] The bottom line in every market is the outcome of the bargaining process, e.g., market price. Although standard axiomatic bargaining theory idealizes the bargaining problem by assuming that individuals are highly rational as they negotiate their desires for various resources, panic and irrational behavior[25] do occur in real-world complex systems including political networks and financial markets. This irrational behavior strongly affects the bargaining process, affects entire complex network systems, and causes bubbles and crashes. Irrespective of market trading rules, players use strategic reasoning and the information available to them when proposing initial bargaining prices.[8, 18, 31, 32, 35, 33, 34] For example, someone selling an apartment needs to know the prices being listed by other sellers of similar apartments in the same neighborhood before they can list an initial bargaining price for their own apartment. The seller takes into account that this initial price will probably be challenged by potential buyers and that bargaining in compeition with other apartment sellers in the neighbor will ensue.

We propose a coupled network model composed of equally sized demand (buyer) and supply (seller) networks, where agents represented as network nodes, cooperate with agents on the opposite side of the market,[18] but compete with agents on the same side of the market, as depicted in Fig. 1. Both demand (buyer) and supply (seller) networks are initialized with a single node, and new nodes together with their prices are added one by one until each network has $N$ nodes. At any given time step and for each demand and supply network the new node is added as follows:

1. With a constant probability $p$, a new node $v_i$ is added to the already existing network structure:

    - node $v_i$ randomly links a node $v_j$, where each $j \in [1, i-1]$ has equal probability $1/(i-1)$ and node $v_i$ is then connected to each neighbor of $v_j$,
    - node $v_i$ sets its market (traded) price $S_i$ as the arithmetic mean of all its neighbors' prices,

    $$S_i = \frac{1}{n_i} \sum_k S_k, \qquad (1)$$

    where $k$ runs over all of $v_i$'s neighbors and the size of $v_i$'s neighborhood is $n_i = n_j + 1$.

2. With a probability $1 - p$ a new node $v_i$ connects to node $e$ in the current network, which has the highest/lowest price $S_e$ in the buyer/seller network, and sets its market price $S_i$ as a percentage increase/decrease $\Delta$ from $S_e$,

$$S_i = S_e \cdot (1 + \Delta). \quad (2)$$

For the limit case $p = 0$, all the nodes form a chain. In the opposite limit case $p = 1$, each new node connects to a node from the existing cluster and all its neighbors, forming a complete graph. These trivial cases together with the case $p = 0.3$ are shown in Fig. 2.

## 2.2 Bargaining Process.

Once both demand and supply networks are generated, the trading between buyers and sellers is initiated. Trading and price dynamics are driven by the inflow of information about an asset, which determines the intrinsic price $S^I$. In contrast to this relatively stable intrinsic price, the dynamics of which will be explained below, the market price is volatile. To generate bubbles and crashes and the feedback mechanism[25] between the trading and price processes, we use network parameters that are time-dependent, not constant, i.e.,

$$p_s(t) = 1 - e^{-\alpha \frac{S(t)}{S^I(t)}}, \ p_d(t) = 1 - e^{-\alpha \frac{S^I(t)}{S(t)}} \quad (3)$$

where $S(t)$ is the last traded price at time $t$. Here $\alpha$ is a parameter controlling overall network clustering.

Note that Eq. (3) implies that when the market price $S(t)$ greatly exceeds the intrinsic value $S^I(t)$, $p_s \to 1$, the competition between supply agents increases, their strategies begin to converge, and their trading prices become increasingly similar (generating the current market price). As a consequence, supply agents become increasingly insecure about their bargaining position. In addition, $p_s \to 1$ implies $p_d \to 0$, which intensifies the market situation because the demand agents now have more strategic options and face less competition. This combination of increasing confidence levels in the demand network and decreasing confidence levels in the supply network increases the probability that there will be an abrupt price crash. This is in agreement with Scheinkman and Xiong who reported that overconfidence generates disagreements among agents regarding asset fundamentals which then causes a significant bubble component in asset prices[36].

Demand and supply agents bargain with each other to reach agreements and execute trades. Here we present a bargaining framework based on the alternating offer strategic model proposed by Rubinstein.[18] Agents are randomly selected, alternating between the demand and supply networks, to make moves. A move is either accepting the best current price offered by the other side or proposing a new price. Each time a trade is executed, the trading agents are removed from their networks and a new one is added in both demand and supply network, thus keeping the number of agents $N$ constant. Agents joining the network follow the proposed network algorithm but are not allowed to connect to the first cluster (the neighborhood of the first node). This is due to the fact that these traders have just bought/sold the asset and believe the price is going to rise/fall, and thus would not go back into their former position immediately. After each trade, there is a probability $\beta$ that the last node from each network will reconnect into the existing network structure following the proposed algorithm, giving the more "fundamental" long-term investors the possibility of changing their position. The utility gained by agent $i$ from trading at price $S$ is quantified by a monotonic utility function $u_i(S)$. Before making a move, agent $i$ evaluates the possible outcome at two steps $\tau = \{0, 1\}$ using a random utility function $U_i(S, \tau)$,

$$U_i(S, 0) = u_i(S), \tag{4}$$

$$U_i(S', 1) = \begin{cases} u_i(S'), & \text{with probability } \lambda_i \\ 0, & \text{with probability } 1 - \lambda_i \end{cases} \tag{5}$$

where $\lambda_i \in [0, 1]$ quantifies "agent confidence," the probability calculated by agent $i$ that his/her offer (price $S'$) will be accepted in the next step at $\tau = 1$. The expected utility of immediately accepting offer $S$ (at $\tau = 0$) is deterministic and equal to the agent's utility function of price $S$, as noted in Eq. (4). The expected utility of proposing a new price $S'$ can be calculated from Eq. (5) as

$$\mathrm{E}[U_i(S', 1)] = u_i(S')\lambda_i. \tag{6}$$

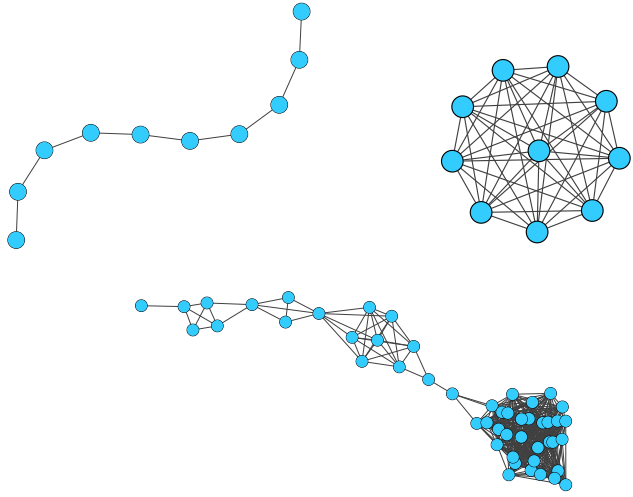

Figure 2: Clustered chain networks of competing agents for $N = 10$, with $p = 0$ (chain) and $p = 1$ (fully connected graph). Clustered chain network of competing agents for $N = 50$ and $p = 0.3$.

Agent $i$ accepts the current offer at $\tau = 0$ if the expected utility at $\tau = 0$ is larger than the expected utility at $\tau = 1$.

Agent $i$ evaluates his/her confidence level $\lambda_i$ based on his/her position in the network, taking into account any available information about trends in the intrinsic price value. Demand and supply agents evaluate their confidence based on the probabilities,

$$\lambda_i^{(s)} = \frac{1}{1 + \frac{n_i^{(s)}}{n_0^{(d)}} \cdot e^{-\gamma r^I(t)}} \qquad \lambda_j^{(d)} = \frac{1}{1 + \frac{n_j^{(d)}}{n_0^{(s)}} \cdot e^{\gamma r^I(t)}} \qquad (7)$$

where $n_i^{(s)}$ and $n_j^{(d)}$ are the neighborhood sizes of supply agent $i$ and demand agent $j$, $n_0^{(s)}$, and $n_0^{(d)}$ is the neighborhood size of the supply and demand agents (the best offers), and $r^I(t)$ is the proportional change in the intrinsic value. Here $\gamma$ is a parameter controlling how much a change in the intrinsic value impacts the market price. When the intrinsic value is approximately constant then the agents evaluate their confidence based on their position in the network only. For supply agent $i$ note that the more competing supply agents $n_i^{(s)}$ rely on similar information and have synchronized trading strategies, and the smaller the number of demand agents $n_0^{(d)}$, the lower the value of $\lambda_i^{(s)}$. More details regarding utility functions $u_i(S)$, new proposed prices $S'$, and agent confidence $\lambda_i$ are given in the Methods section. Note that the competition-cooperation mechanism in trading dynamics in which agents within the same group compete differs from the mechanisms that drive the evolution of cooperation in which agents within the same group cooperate.[37]

## 3 Quantifying the Degree of Market Overpricing

.

In our modelling we were led by the conjecture that market bubbles are caused by complex cooperation-competition bargaining dynamics and that market collapse occurs when the ratio between market price and intrinsic price— close to the Tobin $q$ ratio and its expansions[38, 39]—reaches a critical (tipping) point. To calculate this ratio, which quantifies the degree of market overpricing, we must first estimate the market intrinsic value. Our model for determining the intrinsic value is based on the widely used free cash flow (FCF) model,[40, 41] in which the stock of a company is worth the sum of all of its discounted future free cash flows,[42]

$$S^I(t) = \sum_{j=1}^{\infty} \frac{FCF_{t+j}}{\prod_{k=1}^{j}(1 + WACC_{t+k})}, \qquad (8)$$

and the discount rate is the future weighted average cost of capital (WACC). Using information about past realizations of FCF and WACC and Eq. 8 we determine the intrinsic value $S^I$ at some past time $t - T$. The $S^I$ vaue is approximate because for the $t - T$ time period we know only those FCF and WACC values from $t - T$ up to time $t$ and thus the FCF and WACC values for times longer than $t$ must be estimated. Thus the further back we go ($t \gg T$), the better will be the past price estimate at $t-T$. Because our goal is to estimate the current intrinsic value, we use past $S^I$ values we estimate the current intrinsic value and assume that the growth in cash flows is constant and exponential. For a detailed discussion, see the Methods section. Our model model has two stages, a volatile growth phase that lasts $T$ years followed by a stable "steady state" growth phase. We first examine the steady state growth phase where the ultimate question to be answered is: what are the dynamics that control cash flow?

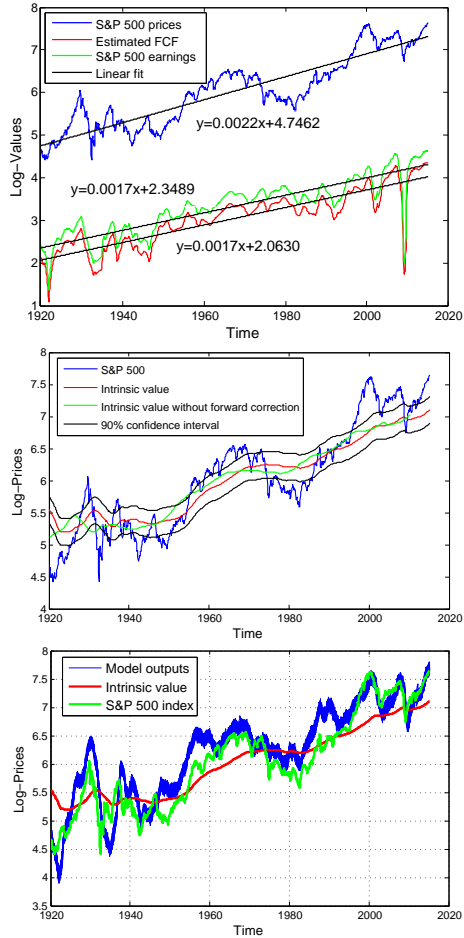

Figure 3: Price dynamics. (a) The S&P 500 prices, earnings, and cash flows, monthly recorded. (b) The S&P 500 index, and the intrinsic value of the S&P 500 index. Both charts are adjusted for inflation. (c) Different time series of model prices together with the intrinsic value of the S&P 500 index from January 1920 to March 2015, and the (market) S&P 500 price index.

To this end, Figure 3(a) shows the time series of Shiller's monthly recorded S&P price, S&P earnings, and S&P cash flows, which represent the behavior of the total US economy for the last 95 years.[43] Note that although the US economy has radically changed over the last century, with traditional industry being replaced by advanced technology, both earning levels and cash flows have on average increased exponentially and the exponential growth rates are approximately constant. This result indicates that the earning dynamics in the US economy during the last two decades have been similar to those during the first two decades of the 20th century. According to Abreu and Brunnermeier[28] market bubbles are sometimes caused when less sophisticated, overly optimistic traders believe that some new technological innovation will guarentee permanently higher growth rates. Figure 3(b) also shows the S&P 500 index $S^M$ over the last 95 years together with the intrinsic value of the S&P 500 index, $S^I$, obtained using Shiller's S&P 500 data from January 1920 until March 2014.[43]

Recalling that in our model trader confidence[25] is quantifed by the ratio between between market price and intrinsic value, Fig. 3(c) shows several realizations of the model market price. These outputs represent all model outputs for this intrinsic price input and no other realizations were discarded. When the market becomes unsustainably overpriced a speculative bubble is created,[25] which in turn makes the agents increasingly uncertain about current and future trading prices. This accumulated collective uncertainity among agents we quantify as the variance of model price, and Fig. 4(a) shows that the variance is at its highest levels just before a market crash when market price greatly exceeds intrinsic value. The accumulated collective uncertainity is further shown in Fig. 4(b) where decreasing confidence in the supply network quantified by increasing the average clustering coefficient is followed by increasing confidence levels in the demand network quantified by decreasing the average clustering coefficient. The results reported in Fig. 4(a)–(b) are in agreement with a complex system phenomenon called critical slowing down[5] characterized by the increase in the variance as the system approaches criticality, which is, in our case, market collapse.

To test how the market and intrinsic prices are related in the long run, Figure 5(a) shows the ratio between the market S&P 500 index and the intrinsic value of the S&P 500 index. Although the intrinsic value of the S&P 500 is not calculated by fitting the market index, the average value of the ratio is close to one (1.04), which is a result that contrasts with the results of a theoretical model that assumes rational agents who do not know the beliefs of other agents and the market price is larger than the intrinsic value.[26] Figure 5(a) further validates our intrinsic price since we find that the peaks of the market-to-intrinsic ratio follow the peaks of the Shiller P/E indicator widely used to estimate the degree of market overpricing. This fact affirms that our $S^I$ represents a reasonable estimation of the intrinsic value of the S&P 500 index. Figure 5(a) further reveals that the periods when the market is underpriced and overpriced occur repeatedly, which agrees, when speaking of companies, with the suggestion that expectations for a company should not be too high or too low[44]. If a company promises investors blue-sky expected outcomes that are not realized, *not only will the share price drop when the market realizes that the company cannot deliver, but it may take years for the company to regain credibility.* To paraphrase Abraham Lincoln, by overestimating the market you can fool some of the people all of the time, but all of the people only some of the time. The final punishment of the market comes in the ensuing period when the market is underpriced.

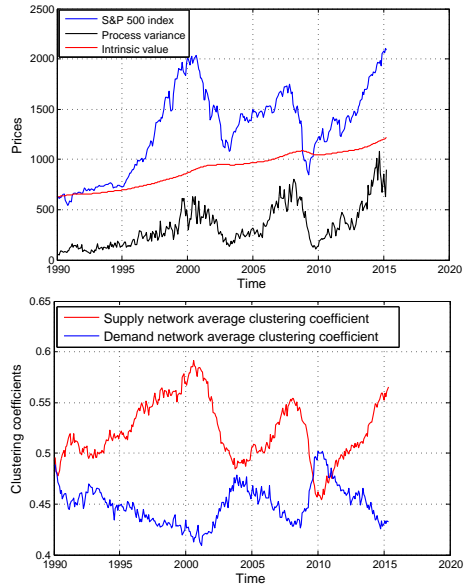

Figure 4: Early-warning indicator. (a) The scaled variance of model price as an early-warning indicator. As the market is getting overpriced, the variance of fluctuations is increasing. (b) Increasing confidence in the demand network is followed with decreasing confidence levels in the supply network.

We confirm the intriguing possibility that two distinct underpriced and overpriced modes are present in the financial complex system. Figure 5(b) shows the probability of the ratio between market price and intrinsic value for the S&P 500 index. We apply a statistical mixture model to reveal the presence of a substructure in the ratio and show that the probability distribution function of the ratio fits the Gaussian mixture model, which resembles the bimodal shape characteristic for small nonlinear systems near a tipping point.[45] Note that we can locate this substructure of the financial system with the underpriced and overpriced modes because we have a model that can estimate the intrinsic price.

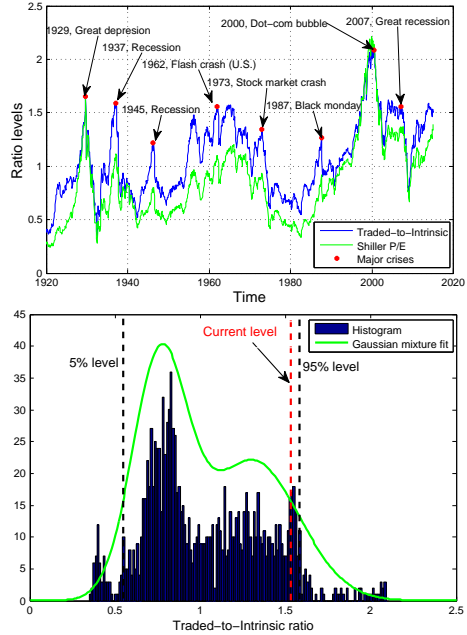

Figure 5: The ratio between the market S&P 500 price index and the estimated intrinsic value of the S&P 500 index. (a) The ratio we compare with the scaled Shiller's P/E (green), the latter obtained by dividing with 20 for the clarity reason. As we note there is a surprising overlap between the peaks. (b) Histogram of the ratio (in blue) resembles a bimodal functional form, characteristic for nonlinear phenomena of small systems near the tipping point. The red vertical line indicates today's value of the ratio, and the fitted Gaussian mixture model with two components at $\mu_1 = 0.8$ and $\mu_2 = 1.3$ is shown in green.

Figure 5(a) suggests that over a long period of time the market price and the intrinsic value should follow each other. We quantify this assumption by investigating the long-term relationship[46] between the S&P 500 index, $S^M(t)$, and the intrinsic value of the S&P 500 index, $S^I(t)$. Motivated by the finding of Campbell and Shiller that dividends and the present discounted value of expected future dividends cointegrate,[46] we next employ the Engle-Granger cointegration test[48, 47] and report the cointegration relationship

$$\log(S^M(t)) - \log(S^I(t)) = 0. \qquad (9)$$

between two series at a 5% confidence level (see Table 2). The test is based on a PP unit root test of regression residuals, with 8 lags included in the Newey-West estimator of the long-run variance (the lag parameter was set to 8 in accordance with the Stock-Watson method[47] $0.75N^{\frac{1}{3}}$ in which $N$ is the number of observations). Details about the test are provided in the Methods section and in Table 2. Intrinsic value and index price may deviate from each other in the short run, but in the long run the intrinsic value catches up to the market realizations. As a consequence of the cointegration, the longer the market is overvalued, the longer we may expect the market to stay in the undervalued mode.

Table 1: Phillips-Perron $Z_t$ unit root tests of S&P 500 intrinsic value and S&P 500 index price; lag length was set to 8 according to the Stock-Watson method; 5% significance critical value equals -8.025.

| Test statistics (intrinsic value) | Test statistics (index price) |
|---|---|
| Levels | |
| 0.2645 | 0.387 |
| First differences | |
| $-20.063$ | $-919.714$ |

Table 2: Engle-Granger cointegration test based on PP unit root test of regression residuals between S&P 500 intrinsic value and S&P 500 index price; lag length was set to 8 according to the Stock-Watson method.

| Test Statistic | Critical values | Significance levels |
|---|---|---|
| $-16.642$ | $-20.5032$ | 1% |
|  | $-14.034$ | 5% |
|  | $-11.213$ | 10% |

## 4  Tipping Points and Hysteresis.

Estimating the tipping points at which a complex system flips from one state to another is a major challenge in network science.[1] Modeling bubbles and crashes in finance, Abreu and Brunnermeier[28] argued that rational arbitrageurs are aware that the market will eventually collapse, but before it happens they want to ride the bubble and generate high returns. They indicate that a bubble bursts when the fraction of speculative traders leaving the market exceeds some threshold, which is the tipping point in their model. In order to estimate tipping points in an arbitrary model, we must first identify the model output and the parameter space within which we search for the tipping point.

We hypothesize that financial crashes occur when a single parameter—the ratio between market price and intrinsic price—reaches a tipping point, and we propose a procedure for estimating tipping points in both network model and U.S. financial market data. For the model we define the network output at time $t$ as the fraction of future times within the time interval $[t + N\Delta t]$ which has a lower price than the current, where $\Delta t$ is the time step and $N$ is an integer. We generate a large number of simulations, and at any $t$ for each price time series $S_{i,t}$ we record the traded-to-intrinsic price ratio $R_{i,t}$ and the information $I_{i,t}$, which indicates what fraction of the future prices are lower than the current (i.e. $S(t+k\Delta t) < S(t), 1 \leq k \leq N$). Combining all pairs $(R_{i,t}, I_{i,t})$, Figure 6(a) shows that the probability of a future price decline versus the ratio $R$ exhibits a hysteresis behavior. The hysteresis is revealed by analyzing the network model when it moves from its underpriced to its overpriced phase and vice versa. For example, when the ratio increases as the market moves into an overpriced phase, along the lower branch (blue line) the probability that the price will drop during $t + N\Delta t$ gradually increases up to some ratio value $R_c$, where the probability of price change exhibits an abrupt change. Prior to reaching the tipping point $R_c$ small increases in ratio induce small increases in the probability of price decline. Upon reaching the tipping point $R_c$ even small changes in $R$ have a huge impact on the probability of price decline and can cause the market to collapse. The tipping point varies beween different simulations—as one would expect for tipping points in the social sciences—and the value $R_c \approx 1.91 \pm 0.19$ is thus an average value. In the upper branch (red line) shown in Fig. 6(a) when the network moves from the tipping point down to some finite ratio, the probability of price decline during $t + N\Delta t$ remains high, $\approx 1$. As the ratio continues to decline, however, the probability drops more rapidly.

As stated above, the tipping point is a stochastic variable and thus not deterministic. In models in which a node's activity is dependent upon the activity of neighboring nodes, quantified by thresholds as in the Watts model,[49, 50] we define thresholds to be fixed numbers and the result is well-defined critical points.[50] However, if we assume that the thresholds are stochastic variables, the critical points will also be stochastic variables.

Figure 6(a) shows that the hysteresis obtained for the network model agrees with the hysteresis shown Fig. 6(b) for the U.S. financial market, represented by the S&P 500. The hysteresis shown in Fig. 6(a) was obtained using a large number of numerical simulations, but the hysteresis in Fig. 6(b) was obtained by analyzing the S&P 500 during the 2000 dot-com crash and the 2007 recession. At any $t$ for the S&P 500—precisely the price time series $S_{i,t}$—we record the ratio $R_t$ of the S&P 500 and the information $I_{i,t}$ indicating what fraction of the future prices are lower than the current $S_{i,t}$ during $t + N\Delta t$. Collecting all of the pairs $(R_t, I_{i,t})$, we calculate for a given $R$ the probability that the price will decline during the time interval $[t + N\Delta t]$. We find that the tipping point at which the U.S. financial system crashes strongly resembles the tipping point we obtain by our model shown in Fig. 6(a).

Because theories with predictive power are highly valued in science and in finance in particular,[51, 52] we calculate future intrinsic values of the S&P 500 by assuming that market earnings and weighted average cost of capital in the future will follow the historical trends. Specifically we use autoregressive models to make predictions of earnings and WACC trends. We further generate a large number of the network realizations, resulting in modeled prices shown in Figure 7(a), and then we count the modeled prices according to their relative change with respect to the maximum or minimum value (which ever occurs earlier) in period from April 2015 to December 2016. Figure 7(a) shows the cumulative frequencies for the cases with the negative change, i.e. loss (in red) and cumulative frequencies with the positive change, i.e. gain (in blue). Figure 7(b) reveals that the probability of S&P 500 declining from its peak for more than 10% is approximately 80%. The same Figure further reveals that there is only 20% chance that the S&P 500 index will grow more than 10% from its minimal value in the same time period. This calculations assumes that the earning trends and WACC obtained from the historical data will continue to hold, at least in the near future, and Figure 3 shows that this assumption is reasonably correct.

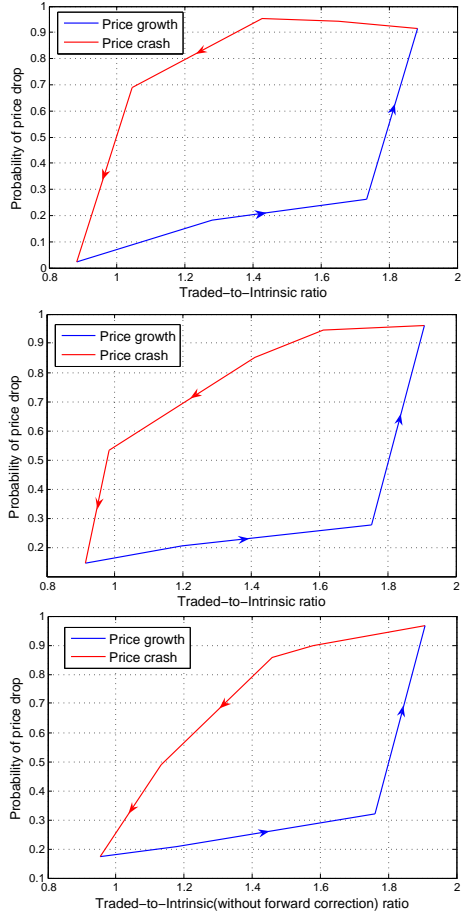

Figure 6: Hysteresis and critical points of financial networks. (a) Model. Probability that the price will drop during $[t, t + N\Delta t]$, where $N = 12$ and $\Delta t = 1$ month, versus the traded-to-intrinsic price ratio, $R$, shows a hysteresis behavior. Approaching a tipping point at $R_c \approx 1.88 \pm 0.12$, the probability of price decline exhibits an abrupt change. (b) S&P 500. Probability that the price will drop during $[t, t + N\Delta t]$ vs. $R$. Hysteresis calculated for the S&P 500 index with a tipping point at $R_c \approx 1.91 \pm 0.19$. (c) S&P 500. Probability that the price will drop during $[t, t+N\Delta t]$ vs. $R$ (where Intrinsic value without forward correction is used). Hysteresis calculated for the S&P 500 index with a tipping point at $R_c \approx 1.9 \pm 0.2$.

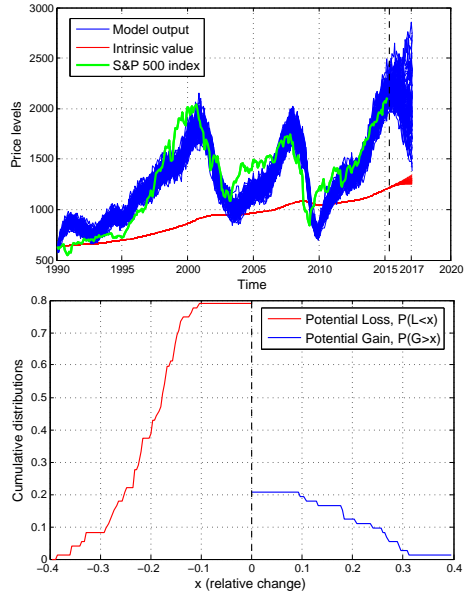

Figure 7: Forecasting power. (a) Different time series of model price (in blue), with the the S&P 500 price index (in green) for the estimated intrinsic values (in red). (b) Cumulative distribution of potential loss $L$, $P(L < x)$ (in red), with cumulative distribution of potential gain $G$, $P(G > x)$ (in blue) in period from April 2015 to December 2016.

## 5 Conclusion.

In this paper we introduce a model that uses complex network science and cooperative game theory to generate trading prices based on fundamental in-formation. This model assumes that agents compete when they use similar strategies. Due to the stochasticity of the network model, using the same input and the same intrinsic value can generate different ouputs, which we can use to analyze trader confidence. The observed market dynamics show that uncertainty among agents is at its highest level just prior to a market crash when the bargaining process produces market values that far exceed the intrinsic values. We hypothesize that the bargaining process around intrinsic values generates bubbles and market crashes.

The model enables us to understand the nature of financial crises and market collapse and thus enhances systemic risk analysis. This will enable us to develop procedures for avoiding global financial meltdowns. In our network model crashes occur when the ratio between market price and intrinsic price, $R$, reaches a tipping point. We use the hysteresis behavior of the network to calculate the tipping point. The probability distribution function of $R$ exhibits a bimodal shape characteristic of small systems near a tipping point. The tipping point before a market crash is not deterministic, and our model can be used to estimate the time interval within which the market will crash. With further insight into agent trading behavior and the effects of different parameter settings, future research will be able to apply our model to the analysis of other important market phenomena (such as bid-ask spread dynamics).

# 6  Methods.

In this section we describe the proposed complex bargaining model in detail and discuss the specifics of the intrinsic price estimation method.

## 6.1  Complex bargaining

Assume two agents ("players") are involved in the sale of a book. Player 1 is selling the book and values it at $60, and player 2 wants to buy the book and values it at $30. The two players are the only agents involved in the transaction, both are rational, and both want the transaction to occur (neither will withdraw). Because there are no other agents who might be competitors and influence the transaction, the two players rationally meet halfway and agree on a price of $45. We can make the situation more complex by adding the assumption that there is a probability $1 - \lambda = 0.8$ that either player will quit the bargaining process before an agreement is reached. In this case each player's anxiety that the other will back out of the transaction gives them greater motivation to reach an agreement. As a result, two equilibrium prices, $35 and $55, emerge. The seller is assured of $35 without risk but must weigh that against getting $55 with a probability of $\lambda = 0.2$. The seller *will accept* any price above $35 but *will not offer* a price below $55. These described bargaining scenarios are rudimentary, and for a more detailed discussion see Nash,[8] Rubinstein,[18] and a number of other contributors.[32, 35]

The agent utility functions are linear functions of the offered price that monotonically increase for supply agents and monotonically decrease for buyer agents, i.e.,

$$u_i^{(s)}(S) = \frac{S - S_{\text{init}}^{(s)}}{S_i^{(s)} - S_{\text{init}}^{(d)}} \; , \; u_j^{(d)}(S) = \frac{S_{\text{init}}^{(s)} - S}{s_{\text{init}}^{(s)} - S_j^{(d)}}, \tag{10}$$

where $S_{\text{init}}^{(s)}$ and $S_{\text{init}}^{(d)}$ are the initial bargaining prices of the supply and demand agents (the best supply and demand offers), respectively. These are reset when a trade occurs and are fixed between trades.

Agents evaluate the utility of accepting $S$, the offer in the current step, using

$$U_i^{(s)}(S, 0) = u_i^{(s)}(S) \; , \; U_j^{(d)}(S, 0) = u_j^{(d)}(S). \tag{11}$$

When evaluating whether to accept an offer in the next bargaining step, an agent must assess the probability that another trading agent will intervene and steal the trading opportunity. For all supply and demand agents, the probability that this "breakdown scenario" ($b$) will occur is $u_i^{(s)}(b) = u_j^{(d)}(b) = 0$. The utility of accepting offer $S$ in the next bargaining step is given by

$$\begin{aligned}\mathrm{E}[U_i^{(s)}(S, 1)] &= u_i^{(s)}(S)\lambda_i^{(s)} + u_i^{(s)}(b)\left(1 - \lambda_i^{(s)}\right) = u_i^{(s)}(x)\lambda_i^{(s)} \\ \mathrm{E}[U_j^{(d)}(S, 1)] &= u_j^{(d)}(xS\lambda_j^{(d)} + u_j^{(d)}(b)\left(1 - \lambda_j^{(d)}\right) = u_j^{(d)}(S)\lambda_j^{(d)}.\end{aligned} \tag{12}$$

Here $\lambda_i^{(s)}$ and $\lambda_j^{(d)}$ are the evaluated probabilities (the "agent confidences") of the supply agent $i$ and demand agent $j$, respectively, poised to trade (or not trade) in the next step. The agents use their neighborhoods to evaluate their confidence following the expression (7). Since external information affects one side positively and the other negatively, supply and demand agents interpret this information reciprocally. In effect, the intrinsic value return $r^I(t)$ tilts the overall confidence evaluations of agents and renders one side more confident and the other less, and is calculated as

$$r^I(t) = \frac{S^I(t) - S^I(t-1)}{S^I(t-1)}. \tag{13}$$

The network parameters used in this study are: $\alpha = 0.2$, $\beta = 0.3$ and $\gamma = 120$. These values were empirically chosen to produce reasonable model outputs.

We use a version of the monotonic concession protocol to simulate agent price movement during the bargaining process [31]. Agents from the supply and demand sides alternate in making their moves. A move can either be accepting the current offer from the other side or proposing a new price. If the offer is accepted and a trade in completed the trading agents are removed from their networks and the agreed-upon price becomes the trading price. An agent accepts the current offer if the expected advantage of accepting it is greater than or equal to the expected advantage of risking that a more attractive price will be agreed upon in the next step. This condition can be expressed using equations (11) and (12) for supply and demand agents,

$$\mathrm{E}\left[U_i^{(s)}\left(S_0^{(d)},0\right)\right] \geq \mathrm{E}\left[U_i^{(s)}\left(S_i^{(s)} \cdot \left(1+\delta_i^{(s)}\right),1\right)\right]$$
$$\mathrm{E}\left[U_j^{(d)}\left(S_0^{(s)},0\right)\right] \geq \mathrm{E}\left[U_j^{(d)}\left(S_j^{(d)} \cdot \left(1+\delta_d^{(j)}\right),1\right)\right]. \qquad (14)$$

If the condition is not met, the agent proposes a new price, $S_i^{(s)} \leftarrow S_i^{(s)} \cdot \left(1+\delta_i^{(s)}\right)$. The multiplicative concession steps $\delta_i^{(s)}$ for supply and demand agents are defined

$$\delta_i^{(s)} = \delta^{(s)}\left(1-\lambda_i^{(s)}\right)\ ,\ \delta_j^{(d)} = \delta^{(d)}\left(1-\lambda_j^{(d)}\right), \qquad (15)$$

where $\delta^{(s)} \leq 0$ and $\delta^{(d)} \geq 0$ are fixed concession step constants for the supply and demand networks, respectively. This means that the magnitude of concession steps made by agents will be determined by the level of agent confidence. Less confident agents offer a larger concession step. After a trade occurs and the trading agents are removed from their networks, a new agent is added to each network, which preserves the orders of both networks. The new nodes are added according to the protocol described above, with the probability $p$ defined in Eq. (3). The parameter values used are $p_c = 0.2$, $\delta^{(d)} = -\delta^{(s)} = 0.005$. The supply and demand networks each have 500 agents and each are initialized with a single node with the initial price equal to the initial intrinsic value. The network steps are $\Delta^{(s)} = 0.01$ and $\Delta^{(d)} = -0.01$ for supply and demand networks, respectively.

## 6.2 Intrinsic price estimation

The free cash flow model (FCFM) is one of the most respected methods for determining the intrinsic value of a company. The model assumes that a company's stock is equal to the sum of all of its future free cash flows (FCF), discounted back to their present value using the future weighted average cost of capital (WACC) as the discount rate (see Eq. (8). FCF is the cash flow generated by the core operations of the business after deducting investments in new capital, and WACC is the rate of return that investors expect to earn from investing in the company. Because investors do not know future values of FCF and WACC, they use estimates based on their own experience, knowledge, and available information. To avoid false precision errors, investors often split their forecast into two periods, (i) a detailed five-year forecast that develops complete balance sheets and income statements with as many links to real variables as possible, and (ii) a simplified forecast for the remaining years. The stock is then traded at a price

$$S(0) = \sum_{j=1}^{T} \frac{FCF_j^E}{\prod_{k=1}^{j}\left(1+WACC_k^E\right)} + \frac{CV_T^E}{\prod_{k=1}^{T}\left(1+WACC_k^E\right)}, \qquad (16)$$

where $FCF_j^E$ is the estimated FCF at the step $j$, $WACC_k^E$ the estimated WACC at time step $k$, and $CV_T^E$ the estimated continuing value at time step $T$. Because the correct FCF and WACC are known for the previous five years—time period $T$—an estimation of a company's intrinsic value at the beginning of that five-year time period, $S^I(-T)$, made today will be more accurate than one made five years ago,

$$S^I(-T) = \sum_{j=-T+1}^{0} \frac{FCF_j}{\prod_{k=-T+1}^{j}(1+WACC_k)} + \qquad (17)$$
$$\sum_{j=1}^{\infty} \frac{FCF_j^E}{\prod_{k=-T+1}^{0}(1+WACC_k)\prod_{k=1}^{j}(1+WACC_k^E)}.$$

Although estimates of FCF and WACC must extend to the indeterminate future, because of the discount effect these estimates will have a decreasing impact on the price. For the sake of simplicity, we assume (i) that the future FCF will grow at a constant growth rate $g$, and (ii) that the future WACC will be constant. Therefore (17) becomes

$$S^I(-T) = \sum_{j=-T+1}^{0} \frac{FCF_j}{\prod_{k=-T+1}^{j}(1+WACC_k)} + \qquad (18)$$
$$\frac{FCF_1}{\prod_{k=-T+1}^{0}(1+WACC_k)(WACC_\infty - g)},$$

where $FCF_1$ is the expected FCF one time step in the future (e.g., one year from now), and $WACC_\infty$ the expected long-term WACC. The goal is to estimate the current intrinsic value of the company now, not its value five years ago. Because over a long period of time the growth rate of the company earnings will be similar to its stock prices, we can estimate the current intrinsic value of the company using

$$S^I(0) = (1+g)^T S^I(-T), \qquad (19)$$

where $g$ is the constant growth rate of earnings. Figure 3(a) shows that this assumption is reasonable, and Fig. 3(b) shows the intrinsic values with and without the forward correction from Eq. (19). Based on the ratio between these two time series, we define the 90% confidence interval, also shown in Fig. 3(b).

Because data for FCF and WACC are usually unavailable, when we implement and test the model we replace these variables with quantities supplied in financial databases. FCF is, in the first approximation,

$$\text{FCF} = \text{E} - \text{NI}, \qquad (20)$$

where E is firm earnings and NI net investments—the increase in invested capital from one year to the next, i.e., the portion of earnings the firm reinvests. The investment rate (IR), the portion of earnings invested back into the business, is

$$\text{IR} = \frac{\text{NI}}{E}. \qquad (21)$$

The return on invested capital (ROIC) is usually defined as the return a company earns on each dollar invested in the business. This is approximately

$$\text{ROIC} = \frac{\text{E}}{\text{Invested Capital}}. \tag{22}$$

Growth ($g$) is usually defined as the rate at which the company's earnings grows each year,

$$g = \frac{\text{NI}}{\text{Invested Capital}}. \tag{23}$$

Note that FCF = E(1− IR) and $g$ = ROIC · IR. Finally,

$$\text{FCF} = \text{E}\left(1 - \frac{g}{\text{ROIC}}\right). \tag{24}$$

In FCFM the discount rate is the weighted average cost of capital. In its simplest form, the weighted average cost of capital is the market-based weighted average of the after-tax cost of debt and cost of equity,

$$\text{WACC} = \frac{\text{D}}{\text{D} + \text{E}}(1 - T_m)k_d + \frac{\text{E}}{\text{D} + \text{E}}k_e, \tag{25}$$

where D/(D + E) is the debt-to-value target level, E/(D + E) the equity-to-value target level, $k_d$ the cost of debt, $T_m$ the marginal tax rate, and $k_e$ the cost of equity.

We use data on the S&P 500 from January 1900 to March 2015 supplied by Robert Shiller when we implement our model. These include data on real (inflation adjusted) earnings, real prices, real long-term interest rates, and S&P500 historical averages (inflation adjusted). The median debt-to-equity ratio for the S&P500 is 19.7%, the median return on invested capital 7%, and the median growth rate of earnings g=1.7%. We use the long-term interest rate as a proxy for the after-tax cost of debt, and the S&P 500 earning yield as a proxy for the cost of equity. Because the time period is $T = 5$ years, at each time step the intrinsic value for the five previous years is determined, and (19) is employed to determine the intrinsic value at the current time step.

We next use the Engle-Granger cointegration test to investigate the long-term relationship [46] between S&P500 intrinsic values and S&P500 index prices. Prior to testing for cointegration, we ensure that both series have the same order of integration. To determine the order of integration, we employ the Phillips-Perron $Z_t$ (PP) unit root test of the null hypothesis that indicates whether the variable has a unit root against a stationary alternative. Table 1 shows the results of PP unit root tests for both levels and the first differences of log-price series. The results imply that both series contain a unit root in levels and thus should be first differentiated to achieve stationarity. We conclude that both series are integrated at the first order, $I(1)$, at a 5% confidence level. To determine whether there is cointegration between the S&P500 intrinsic log-value and the index log-price, we employ the Engle-Granger test,[48, 47] which is based on a PP unit root test of regression residuals, with 8 lags included in the Newey-West estimator of the long-run variance (the lag parameter was set to 8 in accordance with the Stock-Watson method [47] $0.75N^{\frac{1}{3}}$, where $N$ is the number of observations). Table 2 shows that the cointegration between the two series is at a 5% confidence level. Note that intrinsic value and index price can deviate slightly from each other in the short term, but that market forces, government policies, and investor behavior bring them back to a equilibrium in which market realizations and our market expectations converge.

Because theories with predictive power are highly valued in science we calculate future intrinsic values of the S&P 500 by assuming that market earnings and WACC in the future will follow the historical trends. Specifically we use AR(5) model with constant term to make prediction of earnings trend and AR(2) model with constant term to make prediction of WACC trend. Historical earnings and WACC are used for estimation of the models and Akaike information criterion is used for lag order selection in both cases. The fitted models are used in simulation of many possible estimates of the future intrinsic values. We then use these estimates to predict the future market performance.

# Acknowledgments

. We thank J. Cvitanic for useful suggestions.

# References

[1] Scheffer M., Carpenter S., Foley J. A., Folkes C. and Walker B. Catastrophic shifts in ecosystems. *Nature* **413**:591-596 (2001).

[2] Robert M. May, Simon A. Levin, and George Sugihara. Complex systems: Ecology for bankers. *Nature* **451**:893-895 (2008).

[3] Drake J. M. and Griffen, B. D. Early warning signals of extinction in deteriorating environments. *Nature* **456**:456-459 (2010).

[4] Scheffer, M. Foreseeing tipping points *Nature* **467**:411-412 (2010).

[5] Scheffer M. *et al.* Anticipating Critical Transitions. *Science* **338**:344-348 (2012).

[6] Boettiger, C. and Hastings, A. *Nature* **493**:157-158 (2013).

[7] Lontzek, T. S.. Cai, Y., Judd, K. L., and Lenton, T. M. Stochastic integrated assessment of climate tipping points indicates the need for strict climate policy*Nature Climate Change* **5**:441-444 (2015).

[8] Nash, J. F. Jr The Bargaining Problem, *Econometrica* **18**:155-162 (1950).

[9] Perfect Equilibrium in a Bargaining Model. Rubinstein, A. *Econometrica* **50**:97-110 (1982).

[10] Rubinstein, A. and Wolinsky, A. Equilibrium in a Market with Sequential Bargaining, *Econometrica* **53**:295-328 (1985).

[11] Binmore, K. G. and Herrero, M. J. Matching and Bargaining in Dynamic Markets, *Rev. Econ. Stud.* **55**:17-31 (1988).

[12] Gale, D. Bargaining and Competition Part I: Characterization. *Econometrica* **54**:785-806 (1986).

[13] Gale, D. Limit Theorems for Markets with Sequential Bargaining, *J. Econ. Theory* **43**:20-54 (1987).

[14] Manea, M. Bargaining in Stationary Networks, *Amer. Econ. Rev.* **101**:2042-2080 (2011).

[15] West, S. A., Pen, I., and Griffin, A. S. Cooperation and Competition Between Relatives. *Science* **296**:72-75 (1977).

[16] Aguirre, J., Papo, D., and Buldu, J. M. Successful strategies for competing networks. *Nature Physics* **9**:230-234 (2013).

[17] Levin, S. Public goods in relation to competition, cooperation, and spite. *PNAS* **111**:10838-10845 (2014).

[18] Binmore, K., Rubinstein, A., and Wolinsky, A. The Nash Bargaining Solution in Economic Modelling, *The RAND Journal of Economics* **17**:176-188 (1986).

[19] Dobson, I., Carreras, B. A., Lynch, V. E., and Newman, D. E. *Chaos* **17**:026103 (2007).

[20] Noel, Pierre-Andre, Brummitt, C. D., and D'Souza Raissa M., *Phys. Rev. Lett.* **111**:078701 (2013).

[21] Barabasi, A. L. Statistical mechanics of complex networks, *Reviews of Modern Physics* **74(1)**:47-97 (2002).

[22] Brockmann, D. and Helbing, D. The Hidden Geometry of Complex, Network-Driven Contagion Phenomena. *Science* **342**:1337-1342 (2013).

[23] Rosenthal, S. B., Twomey, C. R., Hartnett, A. T., Wu, H. S., and Couzin, I. D. Revealing the hidden networks of interaction in mobile animal groups allows prediction of complex behavioral contagion. *PNAS* **112**:4690-4695 (2015).

[24] Boguna, M., Krioukov, D., and Claffy, K. Navigability of Complex Networks *Nature Physics* **5**:74-80 (2009).

[25] Shiller, R. J. (2000) Irrational Exuberance. *Princeton University Press*.

[26] Allen, F., Morris, S., and Postlewaite, A. Finite Bubbles with Short Sale Constraints and Asyymmetric Information. *Journal of Economic Theory* **61**:206-229 (1993).

[27] De Long, J. B., Shleifer, A. Summers, L. H., and Waldmann, R. J. Noise Trader Risk in Financial Markets. *The Journal of Political Economy* **98**:703-738 (1990).

[28] Abreu, D. and Brunnermeier, M. K. Bubbles and Crashes. *Econometrica* **71**:173-204 (2003).

[29] Vogel, H. L. (2010) Financial Market Bubles and Crashes. *Cambridge Univ Press.*

[30] Scheinkman, Jose A. (2014) Speculation, Trading and Bubbles. *Columbia University Press.*

[31] Rosenschein, J. S. and Zlotkin, G. (2012) Rules of Encounter: Designing Conventions for Automated Negotiation Among Computers, *MIT Press.*

[32] Von Neumann J and Morgenstern, O (1944) Theory of Games and Economic Behavior, *Princeton University Press.*

[33] Maslov, S. Simple model of a limit order-driven market, *Physica A* **278**:571-578 (2000).

[34] Feng, L. Li, B., Podobnik, B., Preis, T., and Stanley, H. E. Linking agent-based models and stochastic models of financial markets, *PNAS* **109**:8388-8393 (2012).

[35] Elliott, M. Inefficiencies in Networked Markets, *American Economic Journal: Microeconomics*, forthcoming.

[36] Scheinkman, Jose A., Xiong, W. Overconfidence and Speculative Bubbles *Journal of Political Economy* **111**:1183-1219 (2003).

[37] Rand, D. G. and Nowak, M. A. Human cooperation. *Trends in Cognitive Sciences* **17**:413-425 (2013).

[38] Brainard, W. C. and Tobin, J. Pitfalls in Financial Model Building. *American Economic Review* **58**: 99-122 (1968).

[39] Blanchard, O., Rhee, C., and Summers, L. The Stock Market, Profit, and Investment. *Quarterly Journal of Economics* **108**: 115-136 (1993).

[40] Gordon, M. J. Dividends, Earnings and Stock Prices. *Review of Economics and Statistics* **41**:99-105 (1959).

[41] Bodie, Z., Kane, A., and Marcus, A. J. Essentials of Investments, (2003) *The McGraw Hill Companies.*

[42] Campbell, J. Y. and Shiller R. J. 1988, Stock Prices, Earnings, and Expected Dividends. *Journal of Finance* **43**:661-676 (1988).

[43] http://www.econ.yale.edu/s̃hiller/data.htm

[44] Koller, T., Goedhart, M. and Wessels, D. Measuring and Managing the value of companies McKinsey & Company, (2005) *John Wiley & Sons, Inc.*

[45] Ding, F. *et al.* Direct molecular dynamics observation of protein folding transition state ensemble. *Biophys. J.* **83**:3525-3532 (2002).

[46] Campbell, J. Y. and Shiller, R. J. Cointegration and Tests of Present Value Models. *The Journal of Political Economy* **95**:1062-1088 (1987).

[47] Hamilton, J. D. (1994) Time Series Analysis. *Princeton, NJ: Princeton University Press.*

[48] Engle, R. F. and Granger C. W. J. Co-Integration and Error-Correction: Representation, Estimation, and Testing. *Econometrica* **55**:251-276 (1987).

[49] Watts, D. J. A simple model of global cascades on random networks. *Proc. Natl Acad. Sci. USA* **99**:5766-5771 (2002).

[50] Majdandzic, A. *et al.* Spontaneous recovery in dynamical networks. *Nature Physics* **10**:34-38 (2014).

[51] Fama, E. F. and French, K. R. The cross-section of expected stock returns, *Journal of Finance* **47**:427-465 (1992).

[52] Balduzzi, P. and Lynch, A. W. Transaction costs and predictability: some utility cost calculations *Journal of Financial Economics* **52**:47-78 (1999).